\documentclass[10pt,twocolumn,twoside,journal]{IEEEtran}
\IEEEoverridecommandlockouts
\usepackage{amsmath,graphicx,amssymb,mathtools,bm}
\usepackage{subfigure}
\usepackage{hyperref}
\usepackage{cite}
\usepackage{amsmath,amssymb,amsfonts}
\usepackage{algorithmic}
\usepackage{graphicx}
\usepackage{textcomp}
\usepackage{xcolor}
\usepackage{verbatim}  
\usepackage{graphicx}  
\usepackage{bm}  
\usepackage{mathrsfs} 
\usepackage{algorithm} 
\usepackage{algorithmic} 
\usepackage{booktabs}
\usepackage{textcomp}  
\usepackage{multirow}  
\usepackage{lettrine}   
\usepackage{graphicx}  
\usepackage{color}  
\usepackage{amsmath}
\usepackage{amssymb}

\def\BibTeX{{\rm B\kern-.05em{\sc i\kern-.025em b}\kern-.08em
		T\kern-.1667em\lower.7ex\hbox{E}\kern-.125emX}}
\begin{document}
	
	\title{Efficient Multi-Beam Training For Terahertz Wireless communications
	}

	\author{Songjie Yang,
		Zhongpei Zhang,~\IEEEmembership{Member,~IEEE},
		Zhenzhen Hu,
		Nuan Song
		and Hao Liu
		\thanks{This work was supported in part by the National Key Research and Development Program of China under Grant 2018YFB1802000, in part by the Guangdong Province Key Project of Science and Technology under Grant 2018B01015001, and in part by NSFC under Grant 61831004. (\textit{Corresponding author:
				Zhongpei~Zhang}.)
		}
		
		\thanks{Songjie Yang and Zhongpei Zhang are with the National Key Laboratory of Science and Technology on Communications, University of Electronic Science and Technology of China, Chengdu 611731, China (e-mail:
			yangsongjie@std.uestc.edu.cn; zhangzp@uestc.edu.cn).
			
			Zhenzhen Hu is with the College of Communication Engineering, Chengdu University of Information Technology, Chengdu 610225, China (e-mail: hzz.uestc@gmail.com).
			
			Nuan Song and Hao Liu are with Bell Labs (China), Shanghai 200120 (e-mail: nuan.song@nokia-sbell.com; Hao.a.liu@nokia-sbell.com)
		}
	}
	\maketitle

	\begin{abstract}
		Although Terahertz communication systems can provide high data rates, it needs high directional beamforming at transmitters and receivers to achieve such rates over a long distance. Therefore, an efficient beam training method is vital to accelerate the link establishment. In this study, we propose a low-complexity beam training scheme of terahertz communication system which uses a low-cost small-scale hybrid architecture to assist a large-scale array for data transmission. The proposed scheme includes two key stages: (\romannumeral1) coarse AoAs/AoDs estimation for beam subset optimization in auxiliary array stage, and (\romannumeral2) accurate AoAs/AoDs estimation by exploiting channel sparsity in data transmission array stage. The analysis shows that the complexity of the scheme is linear with the number of main paths, and thus greatly reduces the complexity of beam training. Simulation results have verified the better performance in spectral efficiency of the
		proposed scheme than that of the related work.
	\end{abstract}
	\section{Introduction}
	\lettrine[lines=2]{W}ith the development of wireless communication technology, terahertz communication technology has attracted more and more attention to meet the increasing data transmission rate. It has become a key technology in the future 6G communication. In order to overcome the serious path loss of terahertz wave, large-scale antenna array is widely used in terahertz communication. 
	However, directional communication complicates the link establishment between the transmitter and the receiver, so a time-consuming beam training procedure must be performed to determine the appropriate transmit and receive directions. Since this will lead to a lot of overhead and waste of network resources, efficient beam training schemes need to be proposed.
	
	Beam training is generally based on codebook for beam sweeping. Through exhaustive search or sequential search, the optimal combinations of analog transmit and receive beams characterizing angle-of-departure (AoD) and angle-of-arrival (AoA) are obtained. However, conventional beam sweeping approaches require a large amount of training. To reduce the training complexity, some low-complexity schemes have been proposed, \cite{train1} developed an algorithm for estimating the channel's AoD and AoA through auxiliary beam pair design with high accuracy and low training overhead. \cite{train2} presented a low complexity multi-level codebook-based beam training algorithm, which is based on level-adaptive antenna selection. \cite{train3} proposed a low complexity primary-auxiliary codebook-based beam training algorithm. In \cite{train4}, the authors proposed to leverage out-of-band information extracted from lower frequency channels to reduce the overhead of establishing a mmWave link. The disadvantage is that the reciprocity between lower frequency channel and mmwave channel is very bad in some cases. By contrast. The auxiliary array and the data transmission array used the same frequency band in \cite{train5}. The auxiliary array of the base station (BS) was used to transmit omnidirectional wave, and the data transmission array of the mobile
	station (MS) estimated AoDs of multiple paths by orthogonal matching pursuit (OMP) algorithm. However, the shortcoming of the scheme is that the performance will be degraded in the case of strong path fading
	in THz band due to the limited coverage of omni-directional wave.
	\begin{figure}
		\includegraphics[width=8.3cm,height=5.3cm]{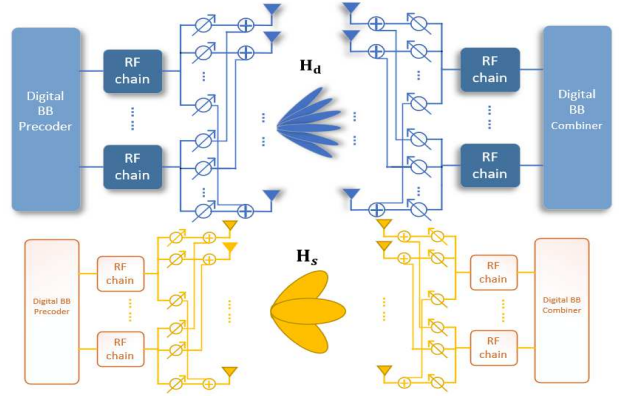}
		\caption{Structure of the data transmission array and the auxiliary array}
	\end{figure}
	
	In this study, we propose a multi-beam training scheme for hybrid beamforming architecture. In the proposed scheme, we consider using a small-scale hybrid antenna array to assist the large-scale hybrid array for data transmission. The auxiliary array and the data transmission array use the same frequency band. In this way, the reciprocity of the two channels is stronger than that between sub-6G and millimeter wave \cite{train4}. Compared with the auxiliary array of fully digital architecture adopted in \cite{train5}, the auxiliary array of our scheme employs a small-scale hybrid structure for less cost consumption. 
	
	The main contributions of our study are as follows:
	\begin{itemize}
		\item We propose an efficient beam training scheme with two stages: auxiliary array stage and data transmission array stage. This method captures the sparse nature of the channel, and greatly reduces the time overhead.
		\item We use a Singular Value Decomposition(SVD)- based method to roughly estimate AoAs/AoDs of the auxiliary array. In this stage, the auxiliary array transmits pilots successively at the transmitter, and the receiver of the auxiliary array receives them with multiple beams simultaneously. We can obtain AoAs/AoDs of the auxiliary array as shown in \ref{proposed}. In order to reduce the consumption of hardware resources, the auxiliary array adopts a small-scale hybrid array.
		\item To compensate for the inaccuracy of AoAs/AoDs estimated by the small-scale auxiliary array, the data transmission array spends a small number of time slots for further beam training to obtain accurate AoAs/AoDs. In this stage, we formulate the beam training of data transmission array as a compressed sensing problem, then we use Compressive Sampling Matching Pursuit (CoSaMP) algorithm to solve the compressed sensing problem. 
	\end{itemize}
	
	\section{System And Channel Model}
	\subsection{System Model}
	We consider two THz system model shown in \figurename{1}. For ease of understanding, we first declare that subscripts $d$ and $s$ represent data array and auxiliary array respectively, and superscripts $t$ and $r$ represent the TX and the RX respectively. Both the BS and MS are equipped with two uniform planar arrays (UPA), two of which are auxiliary arrays with $N_{s}^t$, $N_{s}^r$ antennas and the other two are data transmission arrays with $N_{d}^t$, $N_{d}^r$ antennas. Furthermore, $M_s^t$, $M_s^r$ is the number of RF chains at the transmitter and receiver of the auxiliary array respectively, and $M_d^t$, $M_d^r$ is the number of RF chains at the transmitter and receiver of the data transmission array respectively.
	
	For data transmission array,
	the BS uses $\mathbf{P_d}$ to transmit the signal $\mathbf{S_d}\in\mathbb{C}^{M_d^t\times1}$, the transmitted signal can be expressed as
	\begin{equation}
		\mathbf{X_d}=\rho\mathbf{P_dS_d}
	\end{equation}
	where $\rho$ is a normalization constant to enforce the total radiated power constraint, $\mathbf{P_d}$=$\mathbf{P_d^{RF}P_d^{BB}}$ is the $N_d^t\times M_d^t$ combined BS precoding matrix.
	The signal received by the data transmission array at MS can be written as
	\begin{equation}
		\mathbf{Y_d}=\mathbf{C_d^HH_{d}X_d+C_d^HN_d}
	\end{equation}
	where $\mathbf{C_d}=\mathbf{C_d^{RF}}\mathbf{C_d^{BB}}$ is the $N_d^r\times M_d^r$ MS combiner matrix,  $\mathbf{H_d}\in\mathbb{C}^{N_d^r\times N_d^t}$, $\mathbf{x_d}\in\mathbb{C}^{N_d^t\times1}$ denotes the transmitted signal of the data transmission array and $\mathbf{N_d}\in\mathbb{C}^{N_d^t\times1}$ is the AWGN noise following the distribution $\mathcal{CN}(0,\sigma^2\mathbf{I}_{N_d^t})$.
	
	For the auxiliary array, similarly, the signal received by the auxiliary array at MS can be written as
	\begin{equation}
		\mathbf{Y_s}=\mathbf{C_s^HH_{s}X_s+C_s^HN_s}
	\end{equation}
	
	where
	$\mathbf{X_s}\in\mathbb{C}^{M_s^t\times1}$, $\mathbf{C_s}\in\mathbb{C}^{N_s^r\times M_s^r}$,  $\mathbf{H_s}\in\mathbb{C}^{N_s^r\times N_s^t}$ and $\mathbf{N_s}\in\mathbb{C}^{N_s^t\times1}$ is the AWGN noise following the distribution $\mathcal{CN}(0,\sigma^2\mathbf{I}_{N_s^t})$.
	\subsection{THz Path Gain}
	The path loss of THz channels consists of spreading loss and molecular absorption loss. The path gain of LOS can be expressed as
	\begin{equation}
		\vert\alpha^{LOS}\vert^2=L_{spr}(f,d_0)L_{abs}(f,d_0)
	\end{equation}
	where
	\begin{equation}
		L_{spr}(f,d_0)=(\frac{c}{4\pi fd_0})^2
	\end{equation}
	\begin{equation}
		L_{abs}(f,d_0)=e^{-k_{abs}(f)d_0}
	\end{equation}
	where $f$ is wave frequency, $d_0$ is the distance between the transmitter and the receiver, c is the speed of light and $k_{abs}(f)$ is the absorption coefficient determined by the composition of the transport medium at the molecular level \cite{channel1}.
	
	For the NLOS case, by considering $d_1$ as the distance between the transmitter and the reflector and $d_2$ as the distance between the reflector and the receiver. The path gain of NLOS can be expressed as
	\begin{equation}
		\vert\alpha_l^{NLOS}\vert^2=R_l(f)L_{spr}(f,d_1+d_2)L_{abs}(f,d_1+d_2)
	\end{equation}
	where $R_l(f)$ is the rough surface reflection loss of electromagnetic
	waves at THz band\cite{channel2}. It depends on the material, the shape and the roughness of the surface on which electromagnetic field has been reflected.
	\subsection{Channel Model}
	We first specify the physical channel which characterizes the geometrical structure and limited scattering nature of Thz channels. The physical channel between the data transmission array at the transmitter and the data transmission array at the receiver can be expressed as
	\begin{equation}\label{channel1}
		\mathbf{H_{d}}=\sqrt{\frac{N_{d}^t N_{d}^r}{L}} \sum_{l=1}^L \alpha_l \mathbf{A}_r(\zeta_{\theta}^l,\zeta_{\phi}^l) \mathbf{A}_t^H(\omega_{\theta}^l,\omega_{\phi}^l)  
	\end{equation}
	
	where \emph{L} is the number of paths, $\alpha_l$ is the complex path gain of the \emph{l}-path. Since the typical $N_{d}^{t,y}\times N_{d}^{t,z}$ UPA and $N_{d}^{r,y}\times N_{d}^{r,z}$ UPA are assumed to be equipped in the BS and MS respectively, where $y/z$ represents the azimuth/elevation element in the UPA.  Their array response vectors can be written as 
	\begin{equation}
		\begin{aligned}
			\mathbf{A}_t(\omega_{\theta},\omega_{\phi})=\frac{1}{\sqrt{N_{d}^t}}[1,e^{j2\pi \omega_{\theta}},\cdots,e^{j2\pi (N_{d}^{t}-1)\omega_{\theta}}]^T \\ \otimes [1,e^{j2\pi \omega_{\phi}},\cdots,e^{j2\pi (N_{d}^{t}-1)\omega_{\phi}}]^T 
		\end{aligned}
	\end{equation}
	where \emph{d} is the antenna spacing, $\lambda$ is the wavelength, $\otimes$ denotes the Kronecker product, $\omega_{\theta}=dcos\theta^t/\lambda$  and $\omega_{\phi}=dsin\theta^t sin\phi^t/\lambda$. 
	Similarly, by defining $\zeta_{\theta}=dcos\theta^r/\lambda$ and $\zeta_{\phi}=dsin\theta^r sin\phi^r/\lambda$, we can get the array response of MS
	\begin{equation}
		\begin{aligned}
		\mathbf{A}_r(\zeta_{\theta},\zeta_{\phi})=\frac{1}{\sqrt{N_{d}^r}}[1,e^{j2\pi \zeta_{\theta}},\cdots,e^{j2\pi (N_{d}^{r}-1)\zeta_{\theta}}]^T \\ \otimes [1,e^{j2\pi \zeta_{\phi}},\cdots,e^{j2\pi (N_{d}^{r}-1)\zeta_{\phi}}]^T 
		\end{aligned}
	\end{equation}
	
	Although the physical channel model (\ref{channel1}) has high accuracy, it is difficult to analyze and estimate because of its nonlinear dependence on a large number of parameters ($\alpha_l,\theta_l^r,\phi_l^r,\theta_l^t,\phi_l^t $).  However, due to the limited array aperture, it can be well approximated to linear (in parameter) correspondence, known as a virtual channel model \cite{VCR1,VCR2}.
	The virtual angular domain representation with dictionary matrix of the effective channel $\mathbf{H_{d}}$ can be expressed as
	\begin{equation}\label{VCR}
		\mathbf{H_{d}}=\mathbf{S_r H_a S_t^H}
	\end{equation}
	where $\mathbf{S}_t\in\mathbb{C}^{N_t\times G_t}$,
	\begin{equation}
		\mathbf{S_t}=[\mathbf{A}_t(\omega_\theta^1,\omega_\phi^1),\mathbf{A}_t(\omega_\theta^2,\omega_\phi^2),\cdots,\mathbf{A}_t(\omega_\theta^{G_t},\omega_\phi^{G_t})]
	\end{equation}
	and $\mathbf{S}_r\in\mathbb{C}^{N_r\times G_r}$,
	\begin{equation}
		\mathbf{S_r}=[\mathbf{A}_r(\zeta_\theta^1,\zeta_\phi^1),\mathbf{A}_r(\zeta_\theta^2,\zeta_\phi^2),\cdots,\mathbf{A}_r(\zeta_\theta^{G_r},\zeta_\phi^{G_r})]
	\end{equation}
	
	We assume that the AoAs and AoDs are taken from angular grids of $G_t^y, G_t^z,G_r^y$ and $G_r^z$ points in $(0,\pi]$, respectively. Thus,	$\omega_{\theta}^i$, $\omega_{\phi}^i$, $\zeta_{\theta}^i$ and $\zeta_{\phi}^i$ points in $[-\frac{1}{2},\frac{1}{2})$, respectively.
	\begin{equation}\label{ometh}
		\omega_{\theta}^i=\frac{i-1}{G_t^z}-\frac{1}{2}, \quad i=1,2,\cdots,G_t^z
	\end{equation}
	\begin{equation}\label{omephi}
		\omega_{\phi}^i=\frac{i-1}{G_t^y}-\frac{1}{2},  \quad i=1,2,\cdots,G_t^y
	\end{equation}
	\begin{equation}\label{zath}
		\zeta_{\theta}^i=\frac{i-1}{G_r^z}-\frac{1}{2}, \quad i=1,2,\cdots,G_t^z
	\end{equation}
	\begin{equation}\label{zaphi}
		\zeta_{\phi}^i=\frac{i-1}{G_r^y}-\frac{1}{2}, \quad i=1,2,\cdots,G_t^y
	\end{equation}
	where $G_t=G_t^y\times G_t^z$ and $G_r=G_r^y\times G_r^z$. Besides, $\mathbf{H_a}\in\mathbb{C}^{G_r\times G_t}$ is the angular domain sparse channel matrix with non-zero entries corresponding to the channel path gain, so AoAs/AoDs have one-to-one relationship with the entry index in $\mathbf{H_a}$.
	The $(i$-$j)$th-entry of $\mathbf{H_a}$ for UPA can be written as
	\begin{equation}\label{ha}
		\begin{split}
			\mathbf{H_a}(i,j)\approx \sum_{l=1}^L \alpha_l \emph{f}_{N_{d}^{r,y}}(\frac{m(i)}{G_r^y}-\omega_{\phi}^l)\times \emph{f}_{N_{d}^{r,z}}(\frac{n(i)}{G_r^z}-\omega_{\theta}^l)\\
			\times \emph{f}_{N_{d}^{t,y}}^\ast(\frac{q(j)}{G_t^y}-\zeta_{\phi}^l)\times   \emph{f}_{N_{d}^{t,z}}^\ast(\frac{p(j)}{G_t^z}-\zeta_{\theta}^l)
		\end{split}
	\end{equation}
	
	where $m(i)=\lfloor \frac{i}{G_r^z}\rfloor+1$, $n(i)=mod(\frac{i}{G_r^z})$,  $q(j)=\lfloor \frac{j}{G_t^z}\rfloor+1$, $p(j)=mod(\frac{j}{G_t^z})$ and
	\begin{equation}\label{diric}
		\emph{f}_N(\emph{x})=\frac{1}{N}\sum_{i=0}^{N-1}e^{-j2\pi\emph{x}i}=\frac{1}{N}e^{-j2\pi\emph{x}(N+1)}\frac{sin(\pi N\emph{x})}{sin(\pi\emph{x})}
	\end{equation}
	
	From (\ref{diric}), we note that $\emph{f}_N(\emph{x})$ gets peaky around the origin with increasing $N$. It is obvious that $H_a$ is a sparse matrix with $L$-sparsity.
	
	Similarly, the physical channel between the auxiliary array at the transmitter and the auxiliary array at the receiver can be expressed as

	\begin{equation}\label{channel2}
		\mathbf{H_{s}}=\sqrt{\frac{N_{s}^t N_{s}^r}{L}} \sum_{l=1}^L \alpha_l \tilde{\mathbf{A}}_r(\zeta_{\theta}^l,\zeta_{\phi}^l) \tilde{\mathbf{A}}_t^H(\omega_{\theta}^l,\omega_{\phi}^l)  
	\end{equation}

	It is worth noting that the difference between the data transmission channel model (\ref{channel1}) and the auxiliary channel model (\ref{channel2}) lies in the different number of antennas.
	
	%
	
	\section{Proposed Beam Training}
	\subsection{Auxiliary Array Stage}\label{proposed}
	\begin{algorithm}[!t] 
		\caption{SVD-based AoAs/AoDs Estimation} 
		\label{SVD}      
		\begin{algorithmic}[1] 
			\footnotesize{
				\REQUIRE {Received signals  $\mathbf Y_s$, pilot signals $\mathbf{X_s}$, combining matrix $\mathbf{W_s}$, the number of paths $L$}.
				\ENSURE {Estimated AoAs/AoDs of $L$ paths}.       %
				\STATE{ [$\mathbf{U, \Sigma, V}$]=SVD($\mathbf{Y_s}$)	}.
				\STATE{Select the top $L$ columns ($u_1,u_2,\cdots,u_L$) and ($v_1,v_2,\cdots,v_L$) from $\mathbf{U}$ and $\mathbf{V}$ respectively}.	
				\FOR{$i=1,2,\cdots,L$}
				
				\STATE{$\left(\tilde{\zeta}_{\theta}^i,\tilde{\zeta}_{\phi}^i\right)=\underset{(\zeta_{\theta},\zeta_{\phi})\in\Omega_R}{\mathbf{arg\ max}}$ $u_i^H\mathbf{W_s^H}\mathbf{A}_r(\zeta_{\theta},\zeta_{\phi})$
				}.
				\STATE{$\left(\tilde{\omega}_{\theta}^i,\tilde{\omega}_{\phi}^i\right)=\underset{(\omega_{\theta},\omega_{\phi})\in\Omega_T}{\mathbf{arg\ max}}$ $v_i^H\mathbf{X_s^H}\mathbf{A}_t(\omega_{\theta},\omega_{\phi})$}
				.
				\ENDFOR
			}
		\end{algorithmic}
	\end{algorithm}
	
	In this stage, we use a singular value decomposition(SVD)- based method to roughly estimate AoAs/AoDs of the auxiliary array as shown in algorithm \ref{SVD}.
	
	The channel of auxiliary array in (\ref{channel2}) can  be expressed in matrix form:
	\begin{equation}\label{Hs}
		\mathbf{H_s}=\bm{\mathcal{A}_R}diag(\bm{z})\bm{\mathcal{A}_T}^\mathbf{H}
	\end{equation} 
	
	where
	$\bm{\mathcal{A}_R}$=[$\tilde{\mathbf{A}}_r(\zeta_{\theta}^1,\zeta_{\phi}^1),\tilde{\mathbf{A}}_r(\zeta_{\theta}^2,\zeta_{\phi}^2),\cdots,\tilde{\mathbf{A}}_r(\zeta_{\theta}^L,\zeta_{\phi}^L)$], $\bm{\mathcal{A}_T}$=[$\tilde{\mathbf{A}}_t(\omega_{\theta}^1,\omega_{\phi}^1),\tilde{\mathbf{A}}_t(\omega_{\theta}^2,\omega_{\phi}^2),\cdots,\tilde{\mathbf{A}}_t(\omega_{\theta}^L,\omega_{\phi}^L)$], $\bm{z}$=$\sqrt{\frac{N_{s}^t N_{s}^r}{L}}[\alpha_1,\alpha_2,\cdots,\alpha_L]^T$.
	
	Suppose that the transmiter of auxiliary array sends $P$ pilot sequences $\mathbf{X_s}$= $[\mathbf{x}_1,\mathbf{x}_2,\cdots,\mathbf{x}_{P}]$ to  traverse the $N_X$ beams, where $N_X=PM_s^t$. For each transmit pilot sequence $\mathbf{x}_p(1\leq p\leq P)$, we use $Q$ time slots to obtain an $N_Y$-dimensional received pilot sequence $y_p$, where $N_Y=QM_s^r$. In the $q$-th time slot, we use the combining matrix $\mathbf{W_q}$ to get the received pilot sequence:
	\begin{equation}
		\mathbf{y}_{p,q}=\mathbf{W_q^H}\mathbf{H_sx}_p+\mathbf{n}_{p,q}
	\end{equation} 
	where $\mathbf{W_q}\in\mathbb{C}^{N_s^r\times M_s^r}$. Then we can obtain $	\mathbf{y}_{p}=\mathbf{W_s^H}\mathbf{H_sx}_p+\mathbf{n}_{p}$ by collecting the received pilots in the $M$ time slots. So we have
	\begin{equation}\label{Ys}
		\mathbf{Y_s=W_s^HH_sX_s+N_s}
	\end{equation}
	where $\mathbf{W_s}$=$[\mathbf{w}_1,$$\mathbf{w}_2,\cdots,$$\mathbf{w}_{N_Y}]$$\in\mathbb{C}^{N_s^r\times N_Y}$, $\mathbf{Y_s}$
	=$\mathbf{[y}_1,$ $\mathbf{y}_2,\cdots,$ $\mathbf{y}_{P}]$
	$\in\mathbb{C}^{{N_Y}\times P}$, 
	$\mathbf{X_s}$=$[\mathbf{x}_1$$,\mathbf{x}_2,\cdots$$\mathbf{x}_P]$$\in\mathbb{C}^{N_s^t\times P}$, $\mathbf{N_s}$=$[\mathbf{n}_1$$,\mathbf{n}_2$
	$,\cdots$$\mathbf{n}_P]\in\mathbb{C}^{{N_Y}\times P}$.

	The singular value decomposition of the matrix $\mathbf{Y}$ can be expressed as
	\begin{equation}\label{svdy}
		\mathbf{Y_s=U\Sigma V^H}
	\end{equation}
	
	where $\mathbf{U}$ and $\mathbf{V}$ are unitary matrices, and $\Sigma$=diag$(\sigma_1,\sigma_2,\cdots,\sigma_r)$, $r$ is the rank of $\mathbf{Y_s}$.
	According to (\ref{Hs}), (\ref{Ys}) and (\ref{svdy}), we can get 
	\begin{equation}
		\mathbf{U\Sigma V^H}=\mathbf{W_s^H}\bm{\mathcal{A}_R}diag(\bm{z})\bm{\mathcal{A}}\mathbf{_T^HX_s+N_s}
	\end{equation}
	
	Assume that the noise is small, we can get the largest $L$ singular values and their corresponding singular vectors, and they can be approximately expressed as  
	\begin{equation}
		\sigma_l\approx\vert\bm{z}_l\vert\ \Vert\mathbf{W_s^H}\tilde{\mathbf{A}}_r(\zeta_{\theta}^l,\zeta_{\phi}^l)\Vert_{\tiny{2}} \Vert\mathbf{X_s^H}\tilde{\mathbf{A}}_t(\omega_{\theta}^l,\omega_{\phi}^l)\Vert_2	
	\end{equation}
	\begin{equation}
		u_l\approx \mathbf{W_s^H}\tilde{\mathbf{A}}_r(\zeta_{\theta}^l,\zeta_{\phi}^l)/ \Vert\mathbf{W_s^H}\tilde{\mathbf{A}}_r(\zeta_{\theta}^l,\zeta_{\phi}^l)\Vert_2
	\end{equation}	
	\begin{equation}
		v_l\approx \mathbf{X_s^H}\tilde{\mathbf{A}}_t(\omega_{\theta}^l,\omega_{\phi}^l)/ \Vert\mathbf{X_s^H}\tilde{\mathbf{A}}_t(\omega_{\theta}^l,\omega_{\phi}^l)\Vert_2
	\end{equation}
	
	Therefore, we can search for estimated AoAs/AoDs in the predefined angle domain grids $\Omega_R$/$\Omega_T$ in steps 4 and 5 of Algorithm \ref{SVD}. Where $\Omega_R=\{((i-1)/G_r^z,(j-1)/G_r^y)| i=1,2,\cdots,G_r^z;j=1,2,\cdots,G_r^y\}$ and $\Omega_T=\{((i-1)/G_t^z,(j-1)/G_t^y)| i=1,2,\cdots,G_t^z;j=1,2,\cdots,G_t^y\}$.  
	
	\subsection{Data Transmission Array Stage}
	\begin{algorithm}[!t] 
		\caption{CoSaMP Recovery Algorithm} 
		\label{CoSaMP}      
		\begin{algorithmic} [1]
			\footnotesize{
				\REQUIRE {Received signal $\bm{y}$, measurement matrix $\bm{\Phi}$, sparsity level $L$}.
				\ENSURE {A $L$-sparse approximation $\hat{\theta}$ of the target signal.}     
				\STATE {\textbf{Initialize:} $r_0=\bm{y}$, $\Lambda=\emptyset$, $A=\emptyset$, t=1.}
				\REPEAT
				\STATE{$\textbf{Compute:}$$f=\vert\bm{\Phi}^H r_{t-1}\vert $}.
				\STATE{Select 2$L$ maximum values from $f$, and put the column numbers corresponding to these values into the set $C_0$}.
				
				\STATE{$\Lambda_t=\Lambda_{t-1}\cup C_0, A_t=A_{t-1}\cup \Phi (:,j)$ (for all $j\in C_0$)}.
				
				\STATE {$\hat{\theta}_t=$$\underset{\theta_t}{\mathbf{arg\ min}}\ \Vert \bm{y}-\bm{\Phi} \theta_t\Vert=\left(A_t^HA_t\right)^{-1}A_t^H\bm{y} $.}
				\STATE{Select the largest K values from $ \vert \hat{\theta_t}\vert$ as $ \vert \hat{\theta}_{tK}\vert$ and the corresponding K columns in $A_t$ as $A_{tK}$, update $\Lambda_t=\Lambda_{tK}$.}
				\STATE{$\textbf{Update:}$ $r_t=\bm{y}-A_{tK}\hat{\theta}_{tK}=\bm{y}-A_{tK}(A_{tK}^HA_{tK})^{-1}A_{tK}^H\bm{y}$, $t=t+1$,	$\hat{\theta}=\hat{\theta}_{t-1}$
				}.
				
				\UNTIL{$ t>L\vert\vert r_t=0$
				}

			}
		\end{algorithmic}
	\end{algorithm}
	In this stage, we formulate the beam training of data transmission as compressed sensing problem. After we get 
	$\tilde{\Omega}_{\rm{AOD}}=\{(\tilde{\omega}_{\theta}^1, \tilde{\omega}_{\phi}^1),(\tilde{\omega}_{\theta}^2, \tilde{\omega}_{\phi}^2),\cdots,(\tilde{\omega}_{\theta}^L, \tilde{\omega}_{\phi}^L)$, $ l=1,2,\cdots,L \}$ and
	$\tilde{\Omega}_{\rm{AOA}}=\{(\tilde{\zeta}_{\theta}^1, \tilde{\zeta}_{\phi}^1),(\tilde{\zeta}_{\theta}^2, \tilde{\zeta}_{\phi}^2),\cdots,(\tilde{\zeta}_{\theta}^L, \tilde{\zeta}_{\phi}^L)$ $ l=1,2,\cdots,L \}$, The BS transmits the training signal according to $\tilde{\Omega}_{\rm{AOD}}$ in $L$ time slots. The transmitted data is
	$\mathbf{x}_d^l=P_t\mathbf{A}_t(\tilde{\omega}_{\theta}^l,\tilde{\omega}_{\phi}^l)$ in the $l$-th time slot, $P_t$ is the signal power and is set to $1$. The MS receives the signal with $\mathbf{W}_d^l$, which is composed of $M_d^r$ narrow beams divided by wide beam $\mathbf{A}_r(\tilde{\zeta}_{\theta}^l,\tilde{\zeta}_{\phi}^l)$. Then we can obtain 
	\begin{equation}
		\mathbf{Y}_d^l=(\mathbf{W}_d^l)^\mathbf{H}\mathbf{H}_d\mathbf{x}_d^l+(\mathbf{W}_d^l)^\mathbf{H}\mathbf{N}_d^l
	\end{equation} 
	where $\mathbf{W}_d^l\in \mathbb{C}^{N_d^r\times M_d^r}$. According to (\ref{VCR}), we can get 
	\begin{equation}
		\mathbf{Y}_d^l=(\mathbf{W}_d^l)^\mathbf{H}\mathbf{S_rH_{a}S_t^H}\mathbf{x}_d^l+(\mathbf{W}_d^l)^\mathbf{H}\mathbf{N}_d^l
	\end{equation}
	which is further vectorized to set up the following system
	\begin{equation}\label{r}
		\mathbf{y}_d^l= (\mathbf{W}_d^l)^\mathbf{H}\mathbf{S_r}\mathbf{vec}(\mathbf{H_aS_t^H}\mathbf{x}_d^l)+\mathbf{vec}((\mathbf{W}_d^l)^\mathbf{H}\mathbf{N}_d^l) \\
	\end{equation}
	
	We denote $\mathbf{\Theta}=\mathbf{vec}(\mathbf{H_aS_t^H}\mathbf{x}_d^l)$, the row index of $\mathbf{\Theta}$ corresponds to the AoA in grid, and $\mathbf{\Phi}=(\mathbf{W}_d^l)^\mathbf{H}\mathbf{S_r}$. The measurement matrix $\bm{\Phi}$ is composed of the measurement vectors as its rows and the aim is to reconstruct $\bm{\Theta}$ from $\mathbf{y}_d^l$ and $\bm{\Phi}$. Suppose $\bm{\Theta}$ is sparse, there are theoretical results that have established that $\bm{\Theta}$ in this case can be reconstructed by using either tractable mixed-norm optimization programs\cite{CS1}, greedy algorithms\cite{CS2}, or thresholding algorithms\cite{CS3}. 
	Compared with the traditional OMP algorithm \cite{CS6}, we estimate $\bm{\Phi}$ by Compressive Sampling Matching Pursuit (CoSaMP) \cite{CS5} with the knowledge of $\bm{\Phi}$ and $\bm{\Theta}$. CoSaMP algorithm is based on OMP, but it incorporates several other ideas to accelerate the algorithm and to provide strong guarantees that OMP cannot. Unlike the simplest greedy algorithms, CoSaMP identifies many components during each iteration, which allows the algorithm to run faster for many types of signals. In the OMP algorithm, largest components are calculated and updated per iteration, and then the least square problem is solved. Whereas in the CoSaMP algorithm, all the largest components are identified together, which makes it faster compare to the OMP and it requires only few samples of the vector for reconstruction. Besides, CoSaMP uses the restricted isometry properties of the sampling matrix to ensure that the identification step is  successful. We input $\mathbf{Y}_d^l$ and $\bm{\Phi}$, the output $\hat{\theta}$ is the reconstructed matrix, whose row index corresponds to the AoA in grid. The process of CoSaMP can be seen in Algorithm \ref{CoSaMP}.
	
	\begin{algorithm}[!t] 
		\caption{Proposed Beam Training} 
		\label{beamtrain}      
		\begin{algorithmic} [1]
			\footnotesize{
				\REQUIRE {Coarse AoDs/AoAs set $\tilde{\Omega}_{\rm{AoD}}$/$\tilde{\Omega}_{\rm{AoA}}$}.
				\ENSURE {Accurate AoDs/AoAs set $\hat{\Omega}_{\rm{AoD}}$/$\hat{\Omega}_{\rm{AoA}}$.}     	
				\STATE{RX beam training}.
				\FOR{$i=1,2,\cdots,L$
				}
				\STATE {The BS transmits the signal $\mathbf{A}_t(\tilde{\omega}_{\theta}^l,\tilde{\omega}_{\phi}^l)$, the MS uses $\mathbf{W}_d^l$ to receive data.
				}
				\STATE {According to (31), using algorithm 2 to get reconstructed signal $\hat{\theta}$.
				}
				\STATE {Obtaining $(\hat{\zeta}_{\theta}^l,\hat{\zeta}_{\phi}^l)$ by (16), (17) and (18).
				}
				\ENDFOR

				\STATE{TX beam training}.
				
				\FOR{$i=1,2,\cdots,L$
				}
				\STATE {The MS transmits the signal $\mathbf{A}_r^*(\hat{\zeta}_{\theta}^l,\hat{\zeta}_{\phi}^l)$, the BS uses $\mathbf{F}_d^l$ to receive data.
				}
				\STATE {According to (34), using algorithm 2 to get reconstructed signal $\hat{\theta}$.
				}
				\STATE {Obtaining $(\hat{\omega}_{\theta}^l,\hat{\omega}_{\phi}^l)$ by (14), (15) and (18).
				}
				\ENDFOR
				
			}
		\end{algorithmic}
	\end{algorithm}
	
	We get $\hat{\theta}$, and its nonzero coordinates correspond to the column coordinates of the nonzero terms of $\mathbf{H}_a$. Through $L$ time slots, we can obtain $\hat{\theta}$ of each path. According to (\ref{ometh})(\ref{omephi})(\ref{ha}), we can obtain the AoAs: $\{(\hat{\zeta}_{\theta}^l,\hat{\zeta}_{\phi}^l),l=1,2,\cdots,L\}$. 
	
	After obtaining the AoAs, the MS transmits the signal $\mathbf{\bar{x}}_d^l$=$P_t\mathbf{A}^*_r(\hat{\zeta}_{\theta}^l,\hat{\zeta}_{\phi}^l)$ in the $l$-th time slot, $P_t$ is set to $1$ and $l=1,2,\cdots,L$. We can get 
	\begin{equation}\label{r}
		\mathbf{\bar{y}}_d^l= (\mathbf{F}_d^l)^\mathbf{H}\mathbf{S_t^*}\mathbf{vec}(\mathbf{H_a^TS_r^T}\mathbf{\bar{x}}_d^l)+(\mathbf{F}_d^l)^\mathbf{H}\mathbf{vec}(\mathbf{\bar{N}}_d^l)
	\end{equation}
	where $\mathbf{F}_d^l\in \mathbb{C}^{N_d^t\times M_d^t}$. In the same way, denoting $\mathbf{\Theta}=\mathbf{vec}(\mathbf{H_a^TS_r^T}\mathbf{\bar{x}}_d^l)$, the row index of $\mathbf{\Theta}$ corresponds to the AoD in grid, and $\mathbf{\Phi}= (\mathbf{F}_d^l)^\mathbf{H}\mathbf{S_t^*}$. Next we can obtain the AoD of the $l$-th path by Algorithm 2. The process of obtaining AoAs and AoDs is shown in Algorithm \ref{beamtrain}.
	\subsection{Complexity Analysis}
	In the first stage, the time slots of AoAs and AoDs estimated by the auxiliary array are $QP=N_XN_Y/M_s^tM_s^r$. The time overhead of the second stage is $2L$. However, if  the data transmission array transmits
	data and the auxiliary array estimates the AoAs/AoDs of the next user simultaneously,  the time overhead by the auxiliary array can be ignored. Compared with the $L+1$ time slots for beam training in digital beamforming architecture assisted hybrid beam training (DAH-BT) [5], our proposed scheme costs $L-1$ more time slots in exchange for better performance and less resource consumption.
	\section{Simulation Results}
	In this section, our simulation results are presented in order to demonstrate the performance of our proposed scheme.
	We consider that $f=200GHz$, the LOS component and NLOS component of path gain are given by \cite{channel1},\cite{channel2}. For $d_0 =100m$, $\alpha^{LOS}[dB]\approx-115$, compared with the LOS path, the power of the first-order reflected path is attenuated by more than 10 dB on average. The number of propagation paths $L$ is set to 3. The direct propagation path has a random AoA/AoD, while the other NLOS components have Rayleigh distributed coefficients and uniformly distributed AoAs/AoDs.  In addition, we consider that $N_d^t$, $N_d^r$, $G_t^y$, $G_t^z$, $G_r^y$, $G_r^z$ are all set to 64, and $N_s^t$, $N_s^r$ are set to 16, $M_d^t$, $M_d^r$, $M_s^t$, $M_s^r$ are set to 4, $N_X=N_Y=16$, the antenna element spacing is $d=0.5\lambda$. For the auxiliary array, the beam candidates are selected from a Discrete Fourier Transform (DFT)-based weight vector codebook.

	The performance of proposed scheme can be considered by the spectral effciency $R$.
	\begin{equation}
		R=log_2\vert \mathbf{I}_L+\frac{P_t}{L}\mathbf{R}_n^{-1}\mathbf{C}^H\mathbf{H_d}\mathbf{P}\mathbf{P}^H\mathbf{H}_d^H\mathbf{C}\vert
	\end{equation}
	where $L$ is the number of main paths, $P_t$ is the transmit power, $\mathbf{R}_n$ is the noise covariance matrix, $\mathbf{R}_n$=$\sigma^2\mathbf{C}^H\mathbf{C}$. $\mathbf{P}$ and $\mathbf{C}$ are respectively the precoder and combiner matrices used for communication over $L$ parallel data streams. Thus, $\mathbf{P}$=$[A_t(\hat{\omega}_{\theta}^1,\hat{\omega}_{\phi}^1),A_t(\hat{\omega}_{\theta}^2,\hat{\omega}_{\phi}^2),\cdots,A_t(\hat{\omega}_{\theta}^L,\hat{\omega}_{\phi}^L)]$ and $\mathbf{C}$=$[A_r(\hat{\zeta}_{\theta}^1,\hat{\zeta}_{\phi}^1),A_r(\hat{\zeta}_{\theta}^2,\hat{\zeta}_{\phi}^2),\cdots,A_r(\hat{\zeta}_{\theta}^L,\hat{\zeta}_{\phi}^L)]$.
	\begin{figure}
		\includegraphics[width=8.1cm,height=5.8cm]{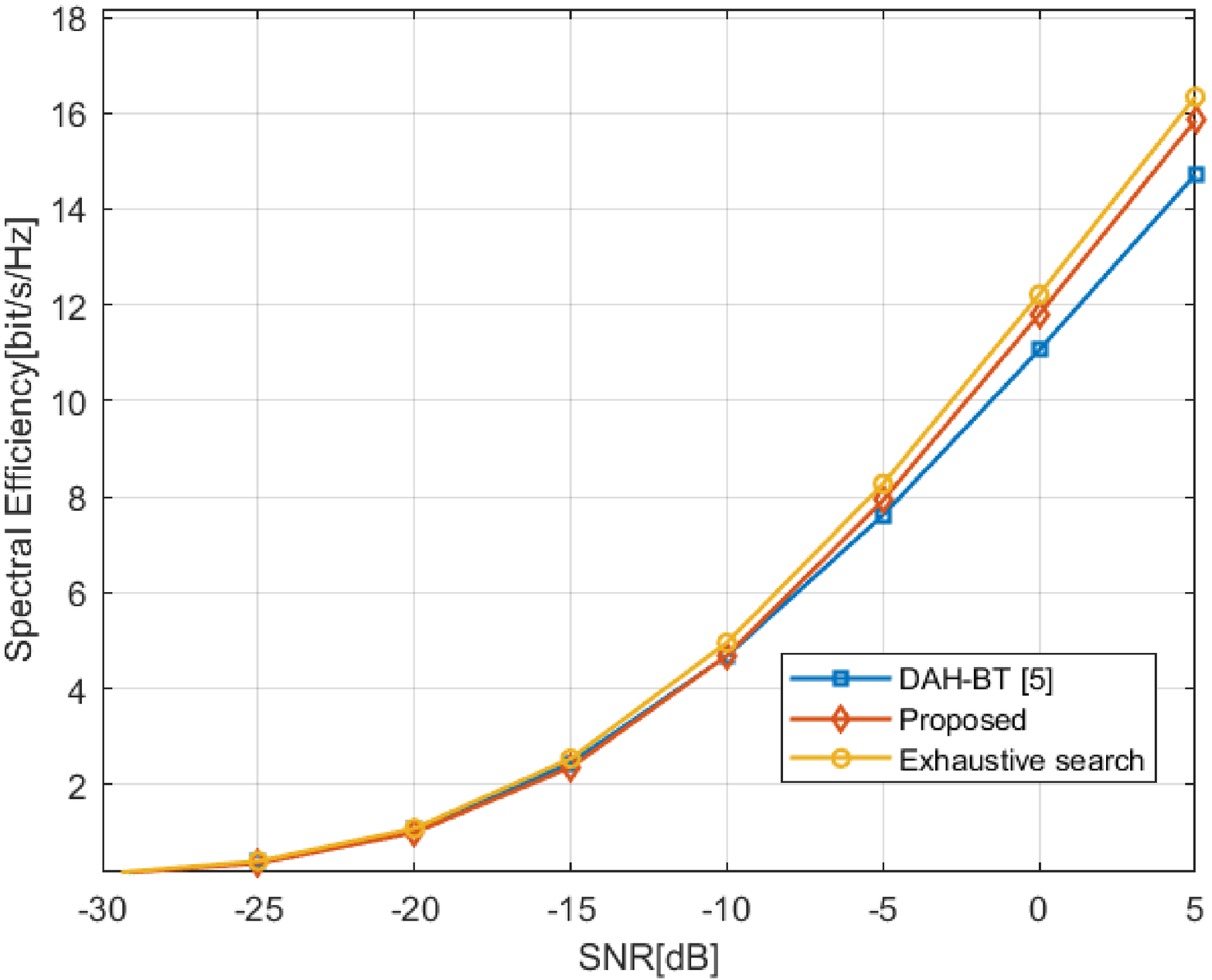}
		\caption{Structure of the data transmission array and the auxiliary array}
	\end{figure}
	
	Fig.2 shows that the spectrum efficiency performance of our scheme is better than that of DAH-BT and is close to that of exhaustive search. Then we define the outage as the event that the spectral efficiency delivered to the MS is below a target value $R_{th}$. Fig.3 shows the outage probalility of the two schemes when $R_{th}=0.1$ and $R_{th}=0.5$ bps/Hz. Obviously, the proposed scheme performs better than DAH-BT in low SNR. 
	\begin{figure}
	\includegraphics[width=8.1cm,height=5.8cm]{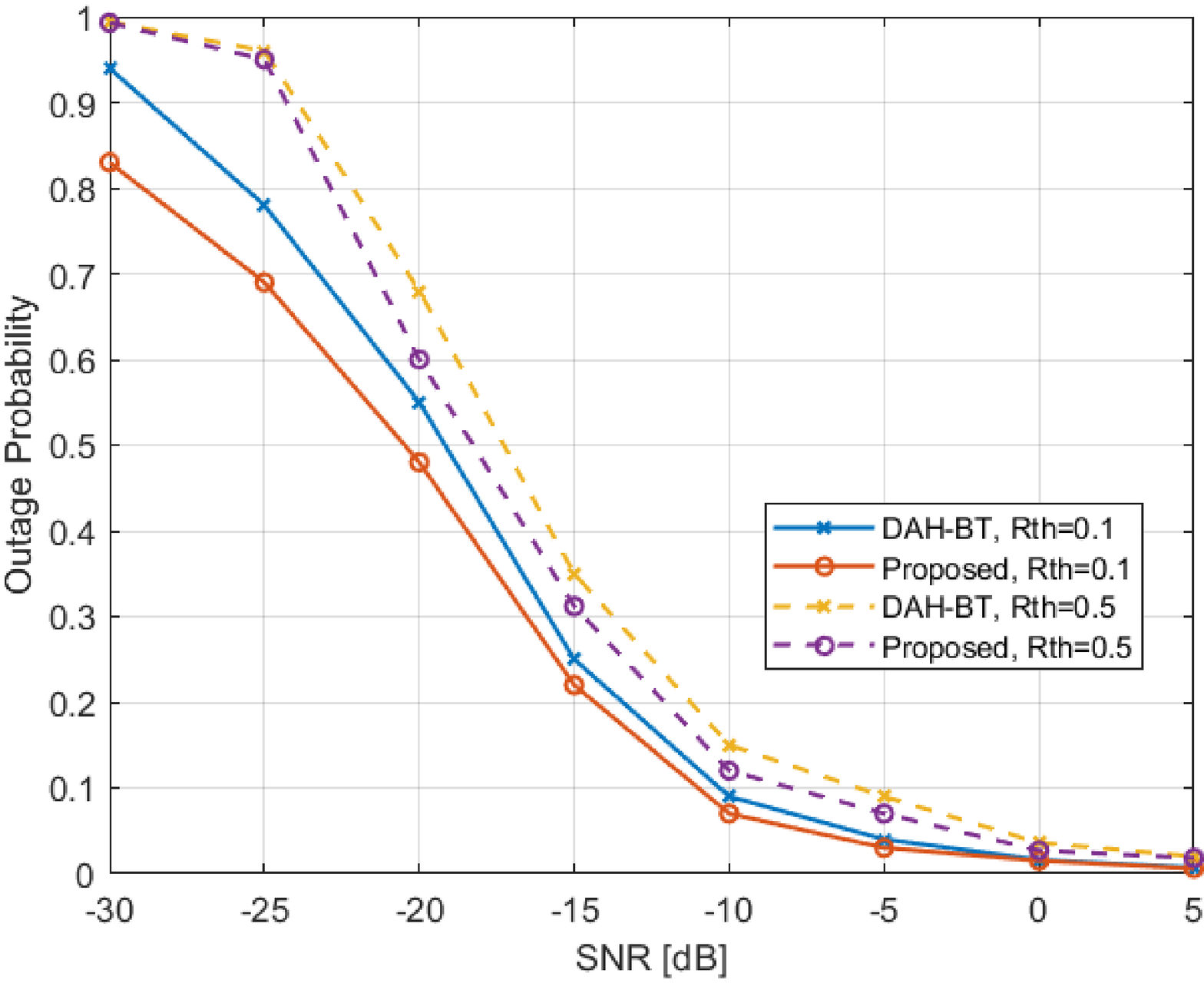}
	\caption{Structure of the data transmission array and the auxiliary array}
\end{figure}
	\section{Conclusions}
	
	In this study, an efficient beam training scheme has been proposed for THz communication system. The scheme includes two stages: the auxiliary array estimates coarse AoAs/AoDs by using a SVD-based method in the first stage, and the data transmission array uses CoSaMP algorithm to estimate accurate AoAs/AoDs by exploiting channel sparsity in the second stage. Analysis and simulation results have shown that our scheme has less cost consumption and better performance than the related method.
	
	%
	\bibliographystyle{IEEEtran}
	\bibliography{reference.bib}

	\vspace{12pt}
	
\end{document}